\documentclass[10pt]{article}

\usepackage{authblk}
\usepackage{amsmath}
\usepackage{booktabs}
\usepackage{bm}
\usepackage{array}
\usepackage{multirow}
\usepackage[colorlinks,citecolor=blue,urlcolor=blue]{hyperref}
\usepackage{graphicx}
\usepackage{bbm}
\usepackage{color}
\usepackage{caption}
\usepackage{epstopdf}
\usepackage{verbatim}
\usepackage[square,sort,comma,numbers]{natbib}
\usepackage{float}

\begin {document}

\title{A semiparametric modeling approach using Bayesian Additive Regression Trees with an application to evaluate heterogeneous treatment effects}

\author[1]{ Bret Zeldow}
\author[2]{Vincent Lo Re III}
\author[3]{Jason Roy}

\affil[1]{\footnotesize Department of Health Care Policy, Harvard Medical School, Boston, MA, USA}
\affil[2]{\footnotesize Department of Medicine, Perelman School of Medicine, Philadelphia, PA, USA}
\affil[3]{\footnotesize Department of Biostatistics, University of Pennsylvania, Philadelphia, PA, USA}

\date{}

%
%
%
%
%
%
%

\maketitle

\begin{abstract}
Bayesian Additive Regression Trees (BART) is a flexible machine learning algorithm capable of capturing nonlinearities between an outcome and covariates and interaction among covariates. We extend BART to a semiparametric regression framework in which the conditional expectation of an outcome is a function of treatment, its effect modifiers, and confounders. The confounders, not of scientific interest, are allowed to have unspecified functional form, while treatment and other covariates that do have scientific importance are given the usual linear form from parametric regression. The result is a Bayesian semiparametric linear regression model where the posterior distribution of the parameters of the linear part can be interpreted as in parametric Bayesian regression. This is useful in situations where a subset of the variables are of substantive interest and the others are nuisance variables that we would like to control for. An example of this occurs in causal modeling with the structural mean model (SMM). Under certain causal assumptions, our method can  be used as a Bayesian SMM. Our methods are demonstrated with simulation studies and an application to dataset involving adults with HIV/Hepatitis C coinfection who newly initiate antiretroviral therapy. The methods are available in an R package semibart.
\end{abstract}


\noindent%
\vfill

%

\section{Introduction}

With an increased emphasis on personalized medicine \cite{precmed} and dynamic treatment regimes \cite{chakraborty2014dynamic}, there is need for statistical methods to evaluate potential effect modifiers. Understanding effect modifiers for a treatment or exposure helps medical personnel, scientists, and politicians better understand mechanisms which can lead to more effective treatment strategies or improved public policies \cite{kan2008season}. In this paper, we introduce a flexible methodology to provide low-dimensional summaries of effect modification when there are a large number of confounders to be controlled for. We implement this through a semiparametric version of Bayesian Additive Regression Trees (BART) \cite{cgm2010} within a generalized linear model (GLM) framework \cite{mccullagh1984generalized}. We model the confounders (the nuisance component) using BART but allow select covariates (treatment, effect modifiers, etc.) to have parametric form. This parametric part can include interactions and typically represents the parameters that are of scientific interest. In practice, these parameters are often estimated in a GLM, but our formulation permits the nuisance component to be nonparametric, increasing efficiency when the outcome process is complex.

Summarizing effect modification can be a difficult task. An introduction to statistical methods for effect modification and interactions is found in \citet{vanderweele2015explanation}. In this paper, we focus on relatively simple summaries of modifiers such as an interaction between the exposure and a covariate. While there are situations where a one dimensional parameter will not adequately capture a complex association, many applications assess effect modification through simple interaction terms \cite{tuut2001smoking, raschenberger2015association}. Our goal is simply to provide a more flexible framework in which to estimate these parameters.

In addition, we show that under some assumptions including no unmeasured confounding, our model serves as a Bayesian implementation of a structural mean model. This fits neatly within the theory laid out in \citet{vansteelandt2003causal} which details estimation of SMMs when g-estimation is not possible, as is the case with a logit or probit link function. We provide an R package \textbf{semibart} to implement our method, available on GitHub (\url{https://www.github.com/zeldow/semibart}). When there is no effect modification or the effect modification is between binary covariates, the framework developed by \citet{hill2012bayesian} can be used. Our model differs from Hill's in that continuous treatments and effect modifiers are more easily interpreted.

We demonstrate our methodology on data from a cohort of HIV/Hepatitis C co-infected individuals who initiate antiretroviral therapy. Certain antiretrovirals are mitochondrial toxic and known to cause liver toxicity, particularly concerning for individuals with Hepatitis C infection. In this paper, we examine two-year death rates in subjects taking antiretroviral regimens with mitochondrial toxic drugs compared to subjects on regimens without those drugs. This analysis is based off a previous analysis that looked at rates of liver decompensation by type of antiretroviral regimen \cite{lo2017risk}.


\section{Background}

We review some semi- and nonparametric methods for predicting an output given inputs.
Let $y$ be an outcome and let $\mathbf{X}$ be predictors of $y$. 
Consider models of the form $g\{E(y | \mathbf{X})\} = \omega(\mathbf{X})$, where $g$ is a known link function and $\omega(\mathbf{x})$ is a function. A special case of this model is parametric linear regression which posits that the outcome has mean $\omega(\mathbf{x}_i; \beta) = \sum^p_{j=1} \beta_j x_{ij}$.
For the remainder of this paper, we relax the assumption that $\omega(\mathbf{x}_i; \beta) = \sum^p_{j=1} \beta_j x_{ij}$, which can yield biased or inefficient estimates if this assumption is far from the truth. 

In practice, the functional form of $\omega(\cdot)$ is unknown and so it is natural to think of it as a random parameter. We assign a prior distribution within an appropriate function space and estimate it with data.  One such prior is the Gaussian process which induces flexibility through its covariance function \cite{rasmussen2006gaussian}. Other options for modeling $\omega(\cdot)$ include the use of basis functions like splines or wavelets and placing prior distributions on the coefficients \cite{muller2015bayesian, eilers1996flexible}. Splines, in particular, have been used extensively in Bayesian nonparametric and semiparametric regression. For example, Biller (2000) presented a semiparametric GLM where one variable is modeled using splines and the remaining variables were part of a parametric linear model \cite{biller2000adaptive}.  Holmes and Mallick (2001) developed a flexible Bayesian piecewise regression using linear splines \cite{holmes2001bayesian}. The approach in Denison et al (1998)  involved piecewise polynomials and was able to approximate  nonlinearities \cite{denison1998automatic}. Biller and Fahrmeir (2001) introduced a varying-coefficient model with B-splines with adaptive knot locations \cite{biller2001bayesian}. 

Alongside these Bayesian methods reside two of the most commonly used procedures to predict an outcome $y$ given covariates $X$: generalized additive models (GAM) \cite{hastie1990generalized} and multivariate adaptive regression splines (MARS) \cite{friedman1991multivariate}. GAM allows each predictor to have its own functional form using splines. However, any interactions between covariates must be specified by the analyst, which can pose difficulties in high-dimensional problems with multi-way interactions. Bayesian versions of GAM based on P-splines exist \cite{brezger2006generalized} but do not have the widespread availability in statistical software that the frequentist version does.  MARS is a fully nonparametric procedure which can automatically detect nonlinearities and interactions through basis functions also based on splines. A Bayesian MARS algorithm has also been developed \cite{denison1998bayesian} but also lacks off-the-shelf software. A third option for nonparametric estimation of $Y$ given $X$ is Bayesian additive regression trees (BART), which like MARS, allows for nonlinear relationships between an outcome and covariates and interactions between covariates, while taking a Bayesian approach to estimation \cite{cgm2010}. 

In this paper we will extend BART to a semi-parametric setting in order to meet our goal of having a parametric form for a subset of variables of interest and a flexible model for the nuisance variables. Before introducing our method, we provide a brief review of BART.

\subsection{Bayesian Additive Regression Trees} \label{sec:BART}

\begin{figure}[!ht]
    \centering
    \includegraphics[width=0.8\textwidth, angle = 0]{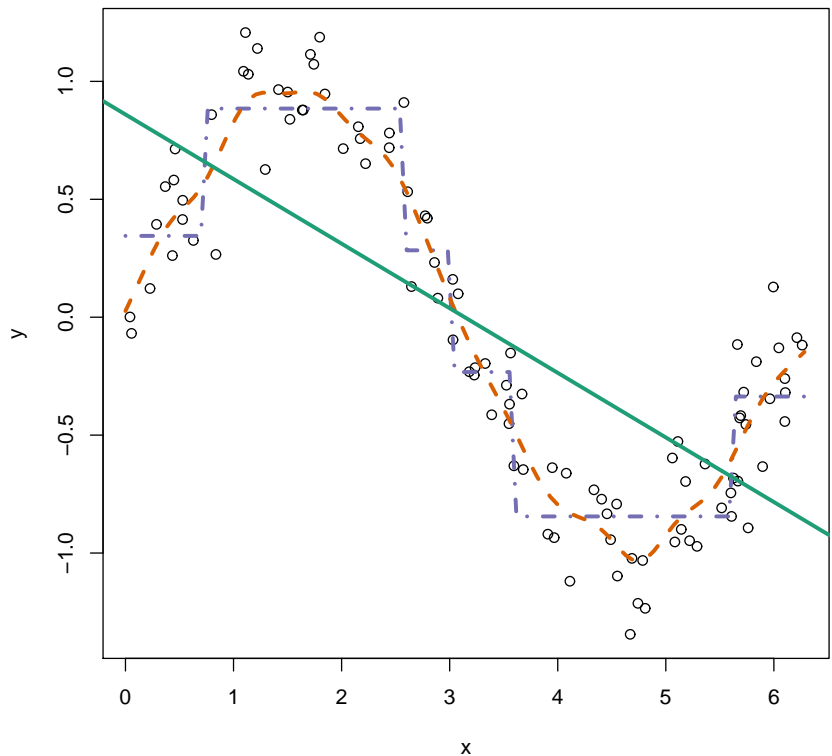}
    \caption{Illustration of a BART fit with a univariate predictor space $x \in [0, 2\pi]$ and mean response $y = \sin(x) + \epsilon$. The solid line is the fit using linear regression, the dashed line is the fit of BART, and the dashed-dotted line is the fit of a single tree.}
  \label{fig:bartex}
\end{figure}

Bayesian additive regression trees (BART) is an algorithm that uses sum-of-trees to predict a binary or continuous outcome given predictors. For continuous outcomes, let $Y = \omega(X) + \epsilon$ where $\epsilon \sim N(0,\sigma^2)$, and $\omega(\cdot)$ is the unknown functional relating the predictors $X$ to the outcome $Y$. For binary $Y$ we use a probit link function so that  $\text{Pr}(Y = 1 | X) = \Phi(\omega(X))$, where $\Phi(\cdot)$ is the distribution function of a standard normal random variable. We write the BART sum-of-trees model as $\omega(x) = \sum_{i=1}^m \omega_i(x; T_i, M_i)$, where each $\omega_i(x)$ is a single tree and $T_i$ and $M_i$ are the parameters that represent the tree structure and end node parameters, respectively. Each individual tree is a sequence of binary decisions based on predictors $X$ which yield predictions of $Y$ within clusters of observations with similar covariate patterns. Typically, the number of trees $m$ is chosen to be large and each tree is restricted to be small through regularization priors, which restricts the influence of any single tree and allows for nonlinearities and interactions that would be not possible with any one tree. We provide an example of a BART fit to a nonlinear mean function $y = \sin(x) + \epsilon$ in Figure~\ref{fig:bartex} over a univariate predictor space $x$ restricted to $[0, 2\pi]$, along with comparision to the fit of a single regression tree and linear regression.

The Markov Chain Monte Carlo (MCMC) algorithm for BART incorporates Bayesian backfitting \cite{hastie2000bayesian}, which we summarize below. Recall that $y_i =  \sum_{j=1}^m \omega_j(x_i; T_j, M_j) + \epsilon_i$ where $\epsilon_i$ is assumed zero-mean normal with unknown variance $\sigma^2$. The algorithm alternates between updates to the error variance $\sigma^2$ and updates to the trees $\omega_j$. To update $\sigma^2$, we find the residuals from the current fit and draw a new value for $\sigma^2$. In \citet{cgm2010} and this paper, we use a conjugate inverse $\chi^2$- distribution for the prior of $\sigma^2$, so drawing a new value is also a draw from an inverse $\chi^2$- distribution. Second, the trees $\omega_j$ are updated one at a time. Starting with $\omega_1$, we compute the residuals of the outcome by subtracting off the fit of the other $m-1$ trees, $\omega_2, \dots, \omega_m$. We then propose a modification for the tree $\omega_1$, which is either accepted or rejected by a Metropolis-Hastings step.  We update the trees $\omega_2, \dots, \omega_m$ in the same fashion. More details are available in the original BART paper \cite{cgm2010}. In the next section, we propose a semiparametric extension of BART, called semi-BART, where a small subset of covariates have linear functional form and the rest are modeled with BART's sum-of-trees.

\section{Semi-BART Model} \label{model}

\subsection{Notation} \label{notation}

Suppose we have $n$ independent observations.  Let $Y$ denote the outcome, which can be binary or continuous. Let $A$ denote treatment, which can be binary or continuous. The remaining covariates we call $\mathbf{X}$. Let $\mathbf{L} = (A, \mathbf{X})$.

\subsection{Semiparametric Generalized Linear Model} \label{slr}

In applied research, it is common for the effects of only a few covariates to be of scientific interest, while a larger number of covariates are needed to address confounding. Our model imposes linearity on just this small subset of covariates, while retaining flexibility in modeling the rest of the covariates whose exact functional form in relation to the outcome may be considered a nuisance. We partition the predictors into two distinct subsets so that $\mathbf{L} = \mathbf{L}_1 \cup \mathbf{L}_2$ and $\mathbf{L}_1 \cap \mathbf{L}_2 = \emptyset$. Here, $\mathbf{L}_1$ represents nuisance covariates that we must control for but is not of primary interest and $\mathbf{L}_2$ represents covariates that do have scientific interest, such as treatment $A$ and its effect modifiers. For continuous $Y$, we write $Y_i = \omega(\mathbf{L}_1) + h(\mathbf{L}_2;\psi) + \epsilon_i$, where $h(\cdot)$ is a linear function of its covariates in $\psi$ (as in linear regression) but $\omega(\cdot)$ is a function with unspecified form. The errors $\epsilon_i$ are iid mean zero and normally distributed with unknown variance $\sigma^2$. More generally, we write $g\left[E(Y | \mathbf{L}_1, \mathbf{L}_2 ) \right] = \omega(\mathbf{L}_1) + h(\mathbf{L}_2; \psi),$ for a known link function $g$. We call this the semi-BART model since we estimate $\omega(\cdot)$ using BART. Note that this implies that if $\mathbf{\mathbf{L}_1} = \mathbf{L}$ and $\mathbf{L}_2 = \emptyset$, we have a nonparametric BART model. On the other hand if $\mathbf{L_1} = \emptyset$ and $\mathbf{L}_2 = \mathbf{L},$ we have a fully parametric regression model. While there is no restriction on the dimensionality of $\mathbf{L_1}$ and $\mathbf{\mathbf{L}_2}$, we assume that that $\mathbf{L_1}$ is large enough such that BART is a reasonable choice of an algorithm and that $\mathbf{\mathbf{L}_2}$ contains only a few covariates that are of particular interest.

\subsection{Special Case: Structural Mean Models} \label{smm}

We now consider a special case of our semiparametric GLM from an observational study with no unmeasured confounders, introducing additional notation specific to this section. As before, the exposure of interest is denoted $A$ and can be either binary or continuous.  The counterfactual $Y^a$ denotes the outcome that would have been observed under exposure $A=a$.  For the special case of binary $A$, each individual has two counterfactual outcomes -- $Y^1$ and $Y^0$ -- but we observe at most one of the two, corresponding to the actual level of exposure received. That is, $Y = A Y^1 + (1 - A) Y^0.$

Robins developed structural nested mean models to adjust for time-varying confounding with a longitudinal exposure \cite{robins1994correcting, robins2000marginal2}.  In the case of a point treatment, structural nested mean models are no longer nested and called structural mean models (SMMs). While time-varying confounding with point treatments is not a concern, SMMs still parameterize a useful causal contrast --  the mean effect of treatment among the treated given the covariates \cite{vansteelandt2003causal, vansteelandt2014structural}. Write this as:
\begin{equation} \label{eq:smm}
g\left\{E\left(Y^a | \mathbf{X} = \mathbf{x}, A= a\right)\right\} - g\left\{E\left(Y^0 | \mathbf{X} = \mathbf{x}, A= a\right)\right\} = h^*(x,a; \psi^*),
\end{equation}
where $g$ is a known link function. 
In this paper, we provide a Bayesian solution to (\ref{eq:smm}). To do so, we impose some restrictions on $h^*(\cdot; \psi^*)$, requiring that under no treatment or when there is no treatment effect the function $h^*(\cdot; \psi^*)$ must equal 0.  That is, $h^*(x,a; \psi^*)$ satisfies $h^*(x,0; \psi^*) = h^*(x,a; 0)  = 0$.  Some examples of $h^*(x, a; \psi^*)$ are $h^*(x,a; \psi^*) = \psi a$ or $h^*(x,a; \psi^*) = (\psi_1 + \psi_2 x_3) a$, when some covariate $x_3$ modifies to the effect of $a$ on $y$. 

While expression (\ref{eq:smm}) cannot be evaluated directly due to the unobserved counterfactuals, two assumptions are needed to identify it with observed data \cite{vansteelandt2014structural}. 

\begin{enumerate}
\item Consistency:  If $A=a$, then $Y^a = Y$;
\item Ignorability: $A \perp Y^0 | X$.
\end{enumerate}
The consistency assumption asserts that we actually get to see an individual's counterfactual corresponding to the exposure received. Ignorability ensures the exposure $A$ and the counterfactual under no treatment $Y^0$ are independent given $X$.  Under these two assumptions together with the parametric assumption of $h^*(\cdot)$, the contrast on the left hand side of (\ref{eq:smm}) is identified, and the SMM from (\ref{eq:smm}) can be rewritten using observed variables as 

\begin{equation} \label{eq:smmbart}
g\left\{E\left(Y | X, A\right)\right\} = \omega(\mathbf{L}_1) + h^*(\mathbf{L}_2; \psi^*), 
\end{equation}
where $\omega(\mathbf{L}_1)$ is unspecified and $h^*(\mathbf{L}_2; \psi^*)$ is a linear function of $\mathbf{L}_2$ \cite{vansteelandt2014structural}. Note that the left hand side of (\ref{eq:smm}) is nonparametrically identified with a third assumption, dropping the parametric assumption of $h^*(\cdot)$. That is,
\begin{enumerate}
\setcounter{enumi}{2}
\item Positivity: $\text{Pr}\left(A = a | \mathbf{X} = \mathbf{x}\right) > 0\ \forall \ x \text{ such that } \text{Pr}(\mathbf{X} = \mathbf{x}) > 0$.
\end{enumerate}
The positivity assumption states that whenever $\mathbf{X} = \mathbf{x}$ has a positive probability of occurring, there is positive probability that an individual is treated. This assumption is violated in situations where treatment is deterministic at certain levels of $\mathbf{X} = \mathbf{x}$.

Let us return briefly to our parametric choice of $h^*(x, a; \psi^*)$, specifically let $h^*(x, a; \psi^*) = (\psi_1 + \psi_2 x_3) a$. Note that the main effect of $x_3$ is not present, guaranteeing that the restriction $h^*(x,0; \psi^*) = h^*(x,a; 0)  = 0$ holds. Through this specification, the function $\omega(\cdot)$ from equation~\ref{eq:smmbart} absorbs the main effect of $x_3$ and any interactions with non-treatment covariates. Practically speaking, we found that modeling a covariate in both $h^*$ and $\omega$ sometimes led to bias and undercoverage of $\psi^*$ in complicated settings. Due to this, we allow the main effect of any effect modifiers to be contained in $h^*$, a choice which comes with a few consequences. First, it imposes linearity on an additional covariate. Second, the restriction $h^*(x,0; \psi^*) = h^*(x,a; 0)  = 0$ no longer holds necessarily. We handle this issue by considering $\omega(x)$ and the term $\psi_3 x_3$ jointly and allowing $(\psi_1 + \psi_2 x_3) a$ to be treated separately. Third, if researchers are interested in quantifying effect modification by $x_3$, they might also be interested in interpreting the main effect and can do so with semi-BART.

Hill (2012) has previously estimated causal effects on the treated using BART \cite{hill2012bayesian}. The methods in that paper correspond to our setting in equation (\ref{eq:smmbart}) where $g$ is the identity link function and $\psi^*$ is a scalar describing only an effect of treatment with no effect modification. We extend this setup to binary outcomes, continuous-valued treatment, or where low-dimensional summaries of effect modification are of interest, particularly with continuous effect modifiers. In settings with continuous outcomes, binary treatment, and no effect modification (or with a binary effect modifier), the methods presented in Hill (2012) are preferred, which we later explore in simulations.

\subsection{Computations}

The algorithm for semi-BART follows the BART algorithm -- briefly reviewed in Section~\ref{sec:BART} -- with an additional step. We solve equation (\ref{eq:smmbart}), where $\omega(\mathbf{L}_1)$ can be written as the sum-of-trees $\sum_{j = 1}^m \omega_j(\mathbf{L}_1; T_j, M_j)$. The parameter $T_j$ contains the structure of the $j^{th}$ tree; for instance, the covariates and rules on which the tree splits. The parameters $M_j$ contain the parameters governing the endnodes of the $j^{th}$ tree. For example, the mean of the $k^{th}$ endnode of the $j^{th}$ tree is assumed to be normally distributed with mean $\mu_{jk}$ and variance $\sigma^2_{jk}$.

For continuous outcomes, we assume independent errors distributed $N(0, \sigma^2)$ with $\sigma^2$ unknown and proceed as follows. Initialize all values including the error variance $\sigma^2$, the parameters $\psi^*$, and the tree structures $T_j$ and $M_j$ for $j = 1, \dots, m$ and iterate through the following steps. First, update the $m$ trees one at a time. Starting with the first tree $\omega_1(\cdot; T_1, M_1)$, calculate the residuals by subtracting the fit of the remaining $m-1$ trees at their current parameter values as well as the fit of the linear part $h^*(\mathbf{L}_2; \psi^*)$. That is for the $i^{th}$ individual, we calculate $y_i^* = y_i - \omega_{-1}(\mathbf{L}_{1i}) - h^*(\mathbf{L}_{2i}; \psi^*)$, where $\omega_{-1}(\mathbf{L}_{1i})$ indicates the fit of the trees except the first tree. As in \citet{cgm2010}, a modification of the tree is now proposed. We can grow the tree (breaking one endnode into two endnodes), prune the tree (collapse two endnodes into one), change a splitting rule (for nonterminal nodes), or swap the rules between two nodes. We accept or reject this modification with a Metropolis-Hastings step given the residuals $\mathbf{y}^* = \{y_1^*, \dots, y_n^*\} $ \cite{cgm1998}. Once we have updated $\omega_1(\cdot; T_1, M_1)$, we update $\omega_2(\cdot; T_2, M_2)$ in the same fashion and continue until all $m$ trees are completed.

Next we update $\psi^*$, given a multivariate normal prior for $\psi^*$ so that $p(\psi^*) \sim \text{MVN}\left(0, \sigma^2_{\psi} \mathbf{I}\right)$, where $\mathbf{I}$ is the identity matrix of appropriate dimension and $\sigma^2_{\psi}$ is large enough so that the prior is diffuse. We calculate the residuals after subtracting off the fit of all $m$ trees so that $y_i^* = y_i - \omega(\mathbf{L}_{1i})$. The posterior for $\psi$ is multivariate normal with covariance $\Sigma_{\psi} = \left[\frac{\mathbf{L}_2^T \mathbf{L}_2}{\sigma^2} + \frac{\mathbf{I}}{\sigma^2_{\psi}}\right]^{-1}$ and mean $\Sigma_{\psi} \left[\frac{\mathbf{L}_2 \mathbf{y}^*}{\sigma^2} + \frac{\bm{\psi}_0}{\sigma^2_{\psi}}\right]$, where $\mathbf{y}^*$ is the $n$-vector of residuals\cite{gelman2014bayesian}.

Lastly, we update the error variance $\sigma^2$. We calculate the residuals given the trees $\omega(\cdot)$ and $\psi^*$ so that $y_i^* = y_i - \omega(\mathbf{L}_{1i}) - h(\mathbf{L}_{2i}; \psi^*)$. With a conjugate scaled inverse $\chi^2$ distribution for $\sigma^2$ (parameters $\nu_0$ and $\lambda_0$), the posterior is an updated scaled inverse $\chi^2$ distribution with parameters $\nu_n = \nu_0 + n$ and $\lambda_n = \nu_0\lambda_0 + <y^*, y^*>$ where $< >$ indicates the dot product. These three steps are repeated until the posterior distributions are well approximated.

Our algorithm for binary outcomes with a probit link uses the underlying latent continuous variable formulation of \citet{albert1993bayesian}, replacing the step in the algorithm that updates the error variance $\sigma^2$. Full details of the BART portion of the algorithm are available in \citet{cgm2010}, whereas our modified code for semi-BART is available at https://www.github.com/zeldow/semibart.

\section{Simulation}

We use simulation to compare the performance of semi-BART to competitor models when estimating the regression coefficient for simulated treatment along with the coefficients for its effect modifiers (main effects and interaction terms with treatment). Our competitors were BART (taken from \citet{hill2012bayesian}), GAM, and linear regression for continuous outcomes and probit regression for binary outcomes.  For all simulations, we generated 500 datasets at sample sizes of $n=500$ and $n=5000$, and we estimated mean bias, 95\% credible/confidence interval coverage, and empirical standard deviation (ESD). For GAM we used the mcgv package in R along with splines (using the s function [the function used to define smooth terms within GAM formulae] with default settings) for continuous covariates\cite{wood2015package}. For BART, we used the bart() function from the BayesTree package in R with default settings\cite{chipman2010bayestree}. The linear regression/probit regression models were fit with the lm and glm functions in R. For semi-BART we used 10,000 MCMC iterations including 2,500 burn-in iterations and $m = 50$ trees.

\subsection{Scenario 1: Continuous outcome with binary treatment and no effect modification}

In the first scenario, we generated data with a continuous outcome, binary treatment, twenty continuous covariates with a block diagonal covariance structure, and four independent binary covariates. The data generating code is available in the appendix. 
 The outcome was generated as independent normal variables with variance one and mean $\mu(a, x) = h(a, x; \psi) + \omega(x)$ where $h(a, x; \psi) = \psi_1 a$ and $\omega(x) = 1 + 2x_1  + \sin (\pi x_2 x_{21} ) -2 \exp( x_{22}x_{24}) + \log | \cos (\frac{\pi}{2} x_3) | - 1.8 \cos (x_4) + 3x_{22}|x_2|^{1.5}$. The parameter $\psi_1$, which encodes the treatment effect, was set to 2.

\begin{table}[!h]
\caption{Simulation results for scenario 1 with continuous outcome and no effect modifiers.}
\label{table1}
\centering
\begin{tabular} { l c  c c c}
\toprule
Method & Parameter  & Bias & Cov. & ESD\\
\midrule
& & \multicolumn{3}{c}{$n = 500$} \\
\multirow{1}{*}{Semi-BART} & $\psi_1$ & -0.02 & 0.96 & 0.153 \\[5pt]
\multirow{1}{*}{GAM}&  $\psi_1$ & -0.02 & 0.94 & 0.371 \\[5pt]
\multirow{1}{*}{BART}& $\psi_1$ & -0.02 & 0.94 & 0.153 \\[5pt]
\multirow{1}{*}{Regression}&  $\psi_1$ & -0.02 & 0.95 & 0.390  \\
\midrule
& & \multicolumn{3}{c}{$n = 5000$} \\
\multirow{1}{*}{Semi-BART} & $\psi_1$ & 0.00 & 0.95 & 0.036 \\[5pt]
\multirow{1}{*}{GAM}&  $\psi_1$ & 0.00 & 0.94 & 0.111 \\[5pt]
\multirow{1}{*}{BART}& $\psi_1$ & 0.00 & 0.92 & 0.037 \\[5pt]
\multirow{1}{*}{Regression}&  $\psi_1$ & 0.01 & 0.94 & 0.119  \\
\bottomrule
\end{tabular}
\end{table}

The results in Table~\ref{table1} show that, at the smaller sample size of $n = 500$, all point estimates are slightly biased in the same direction, and the 95\% coverage probabilities hovered around $95\%$. Notably, the ESD was over half as small for BART-based methods than for GAM or regression. At $n = 5000$ the bias disappeared for all methods, and the discrepancy in ESD between the BART-based methods and non-BART methods remained.

%
%
%

\subsection{Scenario 2: Continuous outcome with binary treatment and continuous effect modifier}

We randomly generated 30 continuous covariates with mean zero from a multivariate normal distribution with an autoregressive(1) covariance $\Sigma$ with $\rho = 0.5$ with the diagonal containing ones. That is,
\[
\Sigma = \begin{pmatrix} 1& \rho & \rho^2 & \rho^3 & \cdots \\
                                                  \rho & 1 & \rho & \rho^2 & \cdots \\
                                                  \rho^2 & \rho & 1 & \rho & \cdots \\
                                                  \rho ^3 & \rho^2 & \rho & 1 & \cdots \\
                                                  \vdots & \vdots & \vdots & \vdots & \ddots

\end{pmatrix}.
\] 

\subsubsection{Part a: simple treatment mechanism and nonlinear mean function}

Given the covariates $x_1 - x_{30}$, the treatment was generated as a Bernoulli random variable with probability $p_a = \text{logit}^{-1}\left(0.1 + 0.2x_1 - \sin(x_3)/3 - 0.1x_{22}\right)$. The outcome was generated as independent random normal variables with variance one and mean $\mu(a, x) = h(a, x; \psi) + \omega(x)$ where $h(a, x) = \psi_1 a + \psi_2 a*x_1 + \psi_3 x_1$ and $\omega(x) = 1 + \sin (\pi x_6 x_{21} ) - \exp( x_{4}x_{5}/5) + \log | \cos (\frac{\pi}{2} x_7) | - 1.8 \cos (x_8) + 0.2x_{10}|x_6|^{1.5}$. The true values for the parameters  are $\psi_1 = 2$, $\psi_2 = -1$, and $\psi_3 = 2$.

Results for these simulations are shown in Table~\ref{tab:ar1}. The estimated parameters are unbiased and have coverage near 95\% for both sample sizes and all estimators. The ESD for all parameters is smaller with semi-BART than it is with GAM or linear regression. This improvement of semi-BART over GAM comes from the fact that covariate interactions are detected in the semi-BART procedure, whereas they must be pre-specified in this implementation of GAM.

\begin{table}[!ht]
\caption{Simulation results for scenario 2a: continuous outcome with simple treatment mechanism and nonlinear outcome. The true parameters are $\psi_1 = 2$, $\psi_2 = -1$, and $\psi_3 = 2$.}
\label{tab:ar1}
\centering
\begin{tabular} { l c  c c c}
\toprule
Method & Parameter  & Bias & Cov. & ESD\\
\midrule
& & \multicolumn{3}{c}{$n = 500$} \\
\multirow{3}{*}{Semi-BART} & $\psi_1$ & -0.01 & 0.94 & 0.123 \\
&  $\psi_2$ & 0.01 & 0.94 & 0.121  \\
&  $\psi_3$ & 0.00 & 0.96 & 0.095 \\[5pt]
\multirow{3}{*}{GAM}&  $\psi_1$ & -0.01 & 0.93 & 0.135 \\
&  $\psi_2$ & 0.01 & 0.94 & 0.127 \\
&  $\psi_3$ & 0.00 & 0.93 & 0.102 \\[5pt]
\multirow{3}{*}{Regression}&  $\psi_1$ & -0.01 & 0.94 & 0.166  \\
&  $\psi_2$ & 0.01 & 0.94 & 0.167 \\
&  $\psi_3$ & 0.00 & 0.94 & 0.127 \\
\midrule
& & \multicolumn{3}{c}{$n = 5000$} \\
\multirow{3}{*}{Semi-BART} & $\psi_1$ & 0.00 & 0.95 & 0.034 \\
&  $\psi_2$ & 0.00 & 0.94 & 0.033  \\
&  $\psi_3$ & 0.00 & 0.96 & 0.023 \\[5pt]
\multirow{3}{*}{GAM}&  $\psi_1$ & 0.00 & 0.94 & 0.038 \\
&  $\psi_2$ & 0.00 & 0.94 & 0.039 \\
&  $\psi_3$ & 0.00 & 0.96 & 0.031 \\[5pt]
\multirow{3}{*}{Regression}&  $\psi_1$ & 0.00 & 0.95 & 0.049  \\
&  $\psi_2$ & 0.00 & 0.95 & 0.049 \\
&  $\psi_3$ & 0.00 & 0.95 & 0.038 \\
\bottomrule
\end{tabular}
\end{table}

\subsubsection{Part b: complex treatment mechanism and complex mean function}

We also performed these simulations with different treatment and outcome data generating functions. Here, given the covariates $x_1 - x_{30}$, the treatment was generated as a Bernoulli random variable with probability $p_a = \text{logit}^{-1}(0.1 + 0.2x_1 - 0.5 x_2 - 0.1x_1x_2 +0.3x_4+0.1x_5+0.7x_4x_5-0.4x_{11}x{22}-0.4x_{10}^2x_{15})$. The outcome was generated as independent random normal variables with variance one and mean $\mu(a, x) = h(a, x; \psi) + \omega(x)$ where $h(a, x) = \psi_1 a + \psi_2 a*x_1 + \psi_3 x_1$ and $\omega(x) = 1 - x_2 +2x_3-1.5x_4-0.5x_5-2x_6+x_3^2-x_6^2+2x_3x_4-x_2x_6+0.5x_5x_6-0.2x_2x_3x_4+x_6x_8x_9-x_7x_{21}x_{24}x_{25}+x_{10}x_{13}x_{14}x_{26}-x_{24}x_{25}^2x_{10}+3x_3x_{16}^2-3x_4x_{17}^2+x_3x_4x_9x_{14}-x_3x_4x_9x_{14}^2+1.5x_{10}x_{21} $. The true values for the parameters are $\psi_1 = 2$, $\psi_2 = -1$, and $\psi_3 = 2$.

Results for these simulations are shown in Table~\ref{tab:ar1extra13}. At $n = 500$, semi-BART yielded biased estimates (average of $-0.07$) for $\psi_3$, the main effect of the effect modifier. On the other hand, GAM and linear regression were unbiased for $\psi_3$ but had varying degrees of bias for the treatment effect $\psi_1$ of $-0.05$ and $0.36$, respectively. Semi-BART had slight undercoverage for all parameters -- 90\% to 92\%. At $n = 5000$, semi-BART was unbiased for $\psi_1$ and $\psi_2$, and the bias of $\psi_3$ attenuated ($-0.12$ down to $-0.03$). For GAM and linear regression, the bias of $\psi_1$ persisted. Coverage rates were all around 95\% save for $\psi_3$ using semi-BART, which was at 89\%. The ESD was notably smaller for semi-BART than the competitors.

\begin{table}[!ht]
\caption{Simulation results for scenario 2b: continuous outcome with complex treatment and outcome functions. The true parameters are $\psi_1 = 2$, $\psi_2 = -1$, and $\psi_3 = 2$.}
\label{tab:ar1extra13}
\centering
\begin{tabular} { l c  c c c}
\toprule
Method & Parameter  & Bias & Cov. & ESD\\
\midrule
& & \multicolumn{3}{c}{$n = 500$} \\
\multirow{3}{*}{Semi-BART} & $\psi_1$ & 0.00 & 0.92 & 0.460 \\
&  $\psi_2$ & -0.01 & 0.92 & 0.459  \\
&  $\psi_3$ & -0.12 & 0.90 & 0.361 \\[5pt]
\multirow{3}{*}{GAM}&  $\psi_1$ & -0.05 & 0.95 & 0.654 \\
&  $\psi_2$ & -0.02 & 0.92 & 0.672 \\
&  $\psi_3$ & 0.00 & 0.93 & 0.520 \\[5pt]
\multirow{3}{*}{Regression}&  $\psi_1$ & 0.36 & 0.92 & 0.731 \\
&  $\psi_2$ & -0.03 & 0.95 & 0.741 \\
&  $\psi_3$ & -0.01 & 0.95 & 0.582 \\
\midrule
& & \multicolumn{3}{c}{$n = 5000$} \\
\multirow{3}{*}{Semi-BART} & $\psi_1$ & 0.00 & 0.96 & 0.081 \\
&  $\psi_2$ & 0.00 & 0.93 & 0.090  \\
&  $\psi_3$ & -0.03 & 0.89 & 0.071 \\[5pt]
\multirow{3}{*}{GAM}&  $\psi_1$ & -0.07 & 0.94 & 0.213 \\
&  $\psi_2$ & 0.00 & 0.95 & 0.200 \\
&  $\psi_3$ & 0.00 & 0.97 & 0.151 \\[5pt]
\multirow{3}{*}{Regression}&  $\psi_1$ & 0.39 & 0.60 & 0.230 \\
&  $\psi_2$ & -0.01 & 0.95 & 0.220 \\
&  $\psi_3$ & -0.01 & 0.96 & 0.166 \\
\bottomrule
\end{tabular}
\end{table}

\subsection{Scenario 3: Binary outcome with binary treatment and continuous effect modifier}

As in scenario 2, we randomly generated 30 continuous covariates with mean zero from a multivariate normal distribution with an autoregressive(1) covariance structure with $\rho = 0.5$ with the diagonal containing ones. The treatment was generated as a Bernoulli random variable with probability $p_a = \text{logit}^{-1}(0.1 + 0.2x_1 - \sin(x_3)/3 - 0.1x_{22})$. The outcome was generated as random Bernoulli variable with probability $p_y(a, x) = \phi\left[h(a, x; \psi) + \omega(x)\right]$ with $h(a, x; \psi) = \psi_1 a + \psi_2 ax_1 + \psi_3 x_1$ and $\omega(x) = 0.1 - \sin(\pi x_6 x_{21}/4) + \exp(x_6/5)x_{11}/4 - 0.12 x_8x_9x_{21} + 0.05x_7x_9x_{10}^2$. The true values for the parameters of interest are $\psi_1 = 0.3$, $\psi_2 = -0.1$, and $\psi_3 = 0.1$.

The results for these simulations are shown in Table~\ref{tab:ar1extra17}. For semi-BART, there is some bias on $\psi_1$ at $n=500$, but this vanishes at $n=5000$. Overall, bias is small and coverage good for both probit regression and semi-BART. Using probit regression is slightly more efficient than semi-BART at $n=500$ (based on ESD) but these differences mostly disappear at $n=5000$.

\begin{table}[!ht]
\caption{Simulation results for Scenario 3: Binary outcome with binary treatment and continuous effect modifier. The true parameter values are $\psi_1 = 0.3$, $\psi_2 = -0.1$, and $\psi_3 = 0.1$}
\label{tab:ar1extra17}
\centering
\begin{tabular} { l c  c c c}
\toprule
Method & Parameter  & Bias & Cov. & ESD\\
\midrule
& & \multicolumn{3}{c}{$n = 500$} \\
\multirow{3}{*}{Semi-BART} & $\psi_1$ & 0.03 & 0.92 & 0.144 \\
&  $\psi_2$ & 0.00 & 0.94 & 0.140  \\
&  $\psi_3$ & 0.00 & 0.93 & 0.106 \\[5pt]
\multirow{3}{*}{Regression}&  $\psi_1$ & -0.01 & 0.93 & 0.131  \\
&  $\psi_2$ & 0.01 & 0.94 & 0.127 \\
&  $\psi_3$ & -0.01 & 0.94 & 0.101 \\
\midrule
& & \multicolumn{3}{c}{$n = 5000$} \\
\multirow{3}{*}{Semi-BART} & $\psi_1$ & 0.00 & 0.94 & 0.039 \\
&  $\psi_2$ & 0.00 & 0.95 & 0.039  \\
&  $\psi_3$ & 0.00 & 0.94 & 0.029 \\[5pt]
\multirow{3}{*}{Regression}&  $\psi_1$ & -0.03 & 0.84 & 0.038  \\
&  $\psi_2$ & 0.01 & 0.93 & 0.036 \\
&  $\psi_3$ & -0.01 & 0.93 & 0.029 \\
\bottomrule
\end{tabular}
\end{table}

\subsection{Scenario 4: Continuous outcome with misspecified mean function}

As before, we randomly generated 30 continuous covariates with mean zero from a multivariate normal distribution with an autoregressive(1) covariance structure. The treatment was generated as a Bernoulli random variable with probability $p_a = \text{logit}^{-1}\left(0.1 + 0.2x_1 - \sin(x_3)/3 - 0.1x_{22}\right)$. The outcome was generated as independent random normal variables with variance one and mean $\mu(a, x) = h(a, x; \psi) + \omega(x)$ with $h(a, x; \psi) = \psi_1 a$ and $\omega(x) = 1 + \sin(\pi x_6x_{21}) - \exp(x_4x_5/5) + \log |\cos(\pi x_7 / 2)| - 1.8 \cos(x_8) + 0.2 x_{10}|x_6|^{1.5} + x_1x_2 - 0.5x_1^2 - \cos(x_1)$. However, we posited the relationship $h(a,x; \psi) = \psi_1 a + \psi_2 a x_1 + \psi_3 x_1$. Since the effect of $x_1$ is actually contained in $\omega(x)$, this is a misspecified model. The true value of $\psi_1$ was 2.

The results for these simulations are shown in Table~\ref{tab:ar1extra18}. All methods have no bias and good coverage for $\psi_1$. There is a slight improvement in terms of ESD for semi-BART compared to its competitors.

\begin{table}[!ht]
\caption{Simulation results for scenario 4: continuous outcome and misspecified mean function. The true value of $\psi_1$ is 2.}
\label{tab:ar1extra18}
\centering
\begin{tabular} { l c  c c c}
\toprule
Method & Parameter  & Bias & Cov. & ESD\\
\midrule
& & \multicolumn{3}{c}{$n = 500$} \\
\multirow{1}{*}{Semi-BART} & $\psi_1$ & 0.01 & 0.95 & 0.143 \\[5pt]
\multirow{1}{*}{GAM}&  $\psi_1$ & 0.00 & 0.92 & 0.153 \\[5pt]
\multirow{1}{*}{Regression}&  $\psi_1$ & 0.01 & 0.96 & 0.177  \\
\midrule
& & \multicolumn{3}{c}{$n = 5000$} \\
\multirow{1}{*}{Semi-BART} & $\psi_1$ & 0.00 & 0.92 & 0.041 \\[5pt]
\multirow{1}{*}{GAM}&  $\psi_1$ & 0.00 & 0.93 & 0.048 \\[5pt]
\multirow{1}{*}{Regression}&  $\psi_1$ & 0.00 & 0.95 & 0.060  \\
\bottomrule
\end{tabular}
\end{table}

\section{Data Application}

To illustrate our method we analyzed data from the Veterans Aging Cohort Study (VACS) in the years 2002 to 2009, which is a cohort of patients being treated at Veterans Affairs facilities in the United States. Our study sample consisted of patients with HIV/Hepatitis C coinfection who were newly initiating antiretrovirals (including at least one nucleoside reverse transcriptase inhibitor [NRTI]) and had at least six months of observations recorded in VACS prior to initiation. Certain NRTIs are known to cause mitochrondial toxicity. These mitochrondial toxic NRTIs (mtNRTIs) include didanosine, stavudine, zidovudine, and zalcitabine \cite{soriano2008antiretroviral}. While these drugs are no longer part of first line HIV treatment regimens, they are still used in resource-limited settings or in salvage regimens \cite{gunthard2016antiretroviral}.

Exposure to mtNRTIs may increase the risk of hepatic injury which in turn may increase the risk of hepatic decompensation and death \cite{scourfield2011value}. The goal of this analysis was to determine if initiating an antiretroviral regimen containing a mtNRTI increased the risk of death versus antiretroviral containing a NRTI that is not a mtNRTI. VACS data contains a number of variables confounding the relationship between mtNRTI use and death including subject demographics, year of antiretroviral initiation, HIV characteristics such as CD4 count and HIV viral load, concomitant medications, and laboratory measures relating to liver function.

One of the covariates included in our analysis is Fibrosis-4 (FIB-4), an index that measures hepatitic fibrosis with higher values indicating larger injury. Specifically FIB-4 $> 3.25$ (no units) indicates advanced hepatic fibrosis.  FIB-4 is calculated as \cite{sterling2006development}:
\[
 \left[\text{age (years)} \times \text{AST (U/L)}\right] / \left[\text{platelet count} (10^9\text{/L}) \times \sqrt{\text{ALT (U/L)}}\right] .
 \]
Here, AST stands for aspartate aminotransferase and ALT for alanine aminotransferase. There is some concern in that mtNRTI use in subjects with high FIB-4 will result in higher risk of liver decompensation and death than in subjects who have lower FIB-4. Thus, we consider FIB-4 as a possible effect modifier of the effect of mtNRTIs on death.

The outcome is a binary indicator of death within a two-year period after the subject initiated antiretroviral therapy. We considered only baseline values for this analysis. There were some missing values among the predictors that were handled through a single imputation. A previous analysis of this data used multiple imputation to handle missing covariates but found that results were very similar across imputations\cite{lo2017risk}. All continuous covariates were centered at interpretable values. For example, age was centered around 50 years and year of study entry was centered at 2005.

In the first analysis we sought to determine the effect of mtNRTI use on death without considering effect modification, and to this extent we fit a Bayesian SMM with a probit link. The estimand can be written as 
\begin{equation}\label{eq:data1}
\Phi^{-1}\left\{E\left(Y^a | \mathbf{X} = \mathbf{x}, A= a\right)\right\} - \Phi^{-1}\left\{E\left(Y^0 | \mathbf{X} = \mathbf{x}, A= a\right)\right\} = \psi a, 
\end{equation}
where $Y$ is the indicator of death, $A$ represents whether mtNRTIs were part of the antiretroviral regimen at baseline ($A = 1$ if mtNRTI were included in the regimen), and $\mathbf{X}$ all other covariates, including FIB-4. In the second and third analysis, we considered FIB-4 to be an effect modifier, once as a continuous covariate and once as a binary indicator which equaled 1 whenever FIB-4 $> 3.25$. This estimand can be written as 
\begin{equation}\label{eq:data2}
\Phi^{-1}\left\{E\left(Y^a | \mathbf{X} = \mathbf{x}, A= a\right)\right\} - \Phi^{-1}\left\{E\left(Y^0 | \mathbf{X} = \mathbf{x}, A= a\right)\right\} = \psi_1 a + \psi_2 a x_1, 
\end{equation}
where $x_1$ corresponds to the appropriate FIB-4 variable.

\begin{figure}[!ht]
    \centering
    \includegraphics[width=0.82\textwidth]{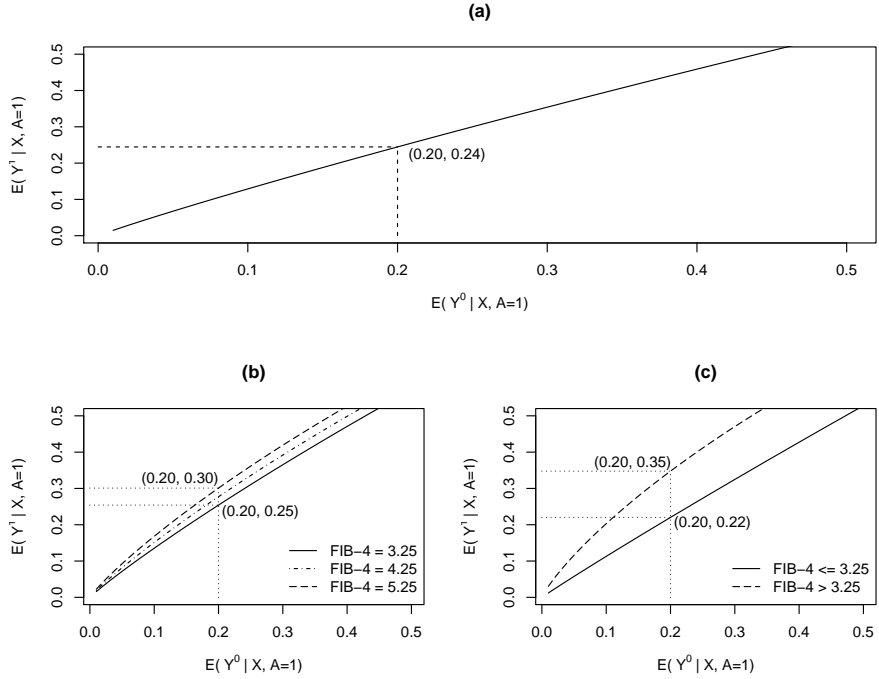}
    \caption{Results of data application showing the effect of having an mtNRTI in an antiretroviral regimen on two-year death. $A = 1$ indicates receipt of an mtNRTI and $A = 0$ indicates no receipt of an mtNRTI. The x-axis represents $E(Y^0 | \mathbf{X}, A = 1)$ which is the mean probability of death if the treated $A = 1$ had in fact been untreated $A = 0$ given $\mathbf{X}$. This quantity is unknown so we consider a spectrum of reasonable values. The y-axis represents $E(Y^1 | \mathbf{X}, A = 1)$ and gives the effect of treatment $A$ on death relative to the x-axis.
(a)  We show effect of mtNRTI on death with no effect modifiers. If $E(Y^0 | \mathbf{X}, A = 1) = 0.20$ then $E(Y^1 | \mathbf{X}, A = 1) = 0.24$. For other values of $E(Y^0 | \mathbf{X}, A = 1)$, identify the value on the x-axis, draw a vertical line until it hits the causal curve, then draw a horizontal line from that point to the y-axis.
(b) We consider the effect modification of mtNRTI on death by continuous FIB-4. Assuming $E(Y^0 | \mathbf{X} = \mathbf{x}, A = 1) = 0.20$,  treatment increases the causal risk of death to 25\% for subjects with FIB-4 = 3.25 (solid line). For subjects with FIB-4 = 4.25 (dotted-dashed line), the causal risk of death increases to 27\%. The causal risk of death for individuals with FIB-4 = 5.25 (dashed line) is 30\%. 
(c) We consider the effect modification of mtNRTI on death by a dichotomized FIB-4. The solid line indicates the causal effect curve when FIB-4 $\leq 3.25$. Assuming $E(Y^0 | \mathbf{X} = \mathbf{x}, A = 1) = 0.20$, we find that treatment increases the mean risk to 22\%. The mean risk of death for individuals with high FIB-4 $> 3.25$ (dashed line) is even higher at 35\%.}
  \label{fig:all}
\end{figure}

The analysis was conducted using $m=50$ trees with 20,000 total iterations (5,000 burn-in). The prior distribution on the $\psi$ parameters were independent $\text{Normal}(0,4^2)$. In the first analysis the mean estimate of the posterior distribution for $\psi$ was 0.15 (95\% credible interval [CI]: -0.02, 0.33). Notably the interval includes 0, but the direction of the point estimate indicates that subjects initiating antiretroviral therapy with an mtNRTI had greater risk of death within 2 years than subjects initiating therapy without an mtNRTI. We can interpret this coefficient in terms of $E\left(Y^0 | \mathbf{X} = \mathbf{x}, A= a\right)$ and $E\left(Y^a | \mathbf{X} = \mathbf{x}, A= a\right)$ from equation (\ref{eq:data1}). Figure~\ref{fig:all}a shows the value of $E(Y^1 | \mathbf{X} = \mathbf{x}, A = 1)$ as a function of $E(Y^0 | \mathbf{X} = \mathbf{x}, A = 1)$ for $\psi = 0.15$. As an example, suppose the unknowable quantity $E(Y^0 | \mathbf{X} = \mathbf{x}, A = 1) = 0.20$. This means that subjects treated with a mtNRTI ($A = 1$) with covariates $\mathbf{X} = \mathbf{x}$ would have had a probability of death of $20\%$ within 2 years had they been untreated ($A = 0$).  However, given $\psi = 0.15$ we see that if $E(Y^0 | \mathbf{X} = \mathbf{x}, A = 1) = 0.20$ then $E(Y^1 | \mathbf{X} = \mathbf{x}, A = 1) = 0.24$, an increase of 4\%. One can examine the change in probability for other base probabilities $E(Y^0 | \mathbf{X} = \mathbf{x}, A = 1)$ by examining the graph in Figure~\ref{fig:all}a.

We conducted a second analysis with FIB-4 as a continuous effect modifier (centered around 3.25) with the same settings as the previous one. This analysis corresponds to the contrast from equation (\ref{eq:data2}). Here, the estimate for the main effect of mtNRTI was $\psi_1 = 0.18$ (0.00, 0.36) and the interaction between mtNRTI use and FIB-4 was $\psi_2 = 0.07$ (0.02, 0.12). The results can be viewed in Figure~\ref{fig:all}b. Again, for illustration, consider the special case where $E(Y^0 | \mathbf{X} = \mathbf{x}, A = 1) = 0.20$. When FIB-4 is 3.25, then $E(Y^1 | \mathbf{X} = \mathbf{x}, A = 1) = 0.25$. However, when FIB-4 is 5.25,  $E(Y^1 | \mathbf{X} = \mathbf{x}, A = 1) = 0.30$.

Finally we did a third analysis with FIB-4 as a binary effect modifier ($> 3.25$ vs. $\leq 3.25$). Here we found that $\psi_1 = 0.07$ (-0.12, 0.26) and $\psi_2 = 0.38$ (0.07, 0.69). These results can be viewed in Figure~\ref{fig:all}c. Here, we see that if $E(Y^0 | \mathbf{X} = \mathbf{x}, A = 1) = 0.20$, then $E(Y^1 | \mathbf{X} = \mathbf{x}, A = 1) = 0.22$ for subjects with FIB-4 $\leq 3.25$ and $E(Y^1 | \mathbf{X} = \mathbf{x}, A = 1) = 0.35$ for subjects with FIB-4 $>3.25$.

\section{Discussion}

We presented a new Bayesian semiparametric model, which can be implemented with an R package semibart that is available on the author's GitHub page (https://github.com/zeldow/semibart). Our model allows for flexible estimation of the nuisance component while being parametric for covariates that are of immediate scientific interest, which provides a viable and intuitive alternative to fully parametric regression. Under some causal assumptions, this model can as be interpreted as a SMM, which also provides the first fully Bayesian SMM. This is particularly useful in the case of binary outcomes where g-estimation is not possible.  \citet{vansteelandt2003causal} provided approaches for estimating SMMs with binary outcomes in frequentist settings; our method is consistent with their suggestions but incorporates the added flexibility of BART.

The primary limitation of our method in the causal setting is that semi-BART does not currently accommodate instrumental variables or longitudinal treatment measures, which are frequently used components of structural nested models. In simulations with binary outcomes, we also found little difference in our estimates using semi-BART versus probit regression. Although it is reassuring that semi-BART works as well as parametric regression, we aim to better understand the reasons why we are seeing equivalent -- rather than superior -- performance of semi-BART (versus probit regression) for binary outcomes. Lastly, we aim to extend semi-BART to handle common link functions such as logit and log.

We imagine that our work can benefit a number of different types of researchers. In particular, we hope that semi-BART can be a viable alternative to the researcher who uses linear regression as the default statistical method in applications. Second, we also hope to convince researchers who prefer flexible machine learning algorithms, such as BART, but need interpretable coefficients such as a treatment effect and its modifiers that semi-BART is a dependable option.

\section*{Supplementary Material}
\appendix


\section{Simulation code - data generation}

\subsection{Scenario 1}

\begin{verbatim}
library(mvnfast)

p1 <- 20 # no. of continuous covariates
p2 <- 5 # no. of binary

makedata <- function(nsim, n, p1, p2, sd) {
  dat4 <- array(0, dim = c(nsim, n, p1 + p2 + 1) )
  for (k in 1:nsim) {
    cov <- matrix(0, p1, p1)
    diag(cov) <- rep(1, p1)

    for(i in 1:5) {
      for(j in 1:5) {
        if (i != j) cov[i,j] <- cov[j, i] <- 0.20
      }
    }
    for(i in 6:10) {
      for(j in 6:10) {
        if (i != j) cov[i,j] <- cov[j, i] <- 0.15
      }
    }
    for(i in 11:15) {
      for(j in 11:15) {
        if (i != j) cov[i,j] <- cov[j, i] <- 0.10
      }
    }
    for(i in 16:20) {
      for(j in 16:20) {
        if (i != j) cov[i,j] <- cov[j, i] <- 0.05
      }
    }

    mu <- c(rep(2.0, 5), rep(1.5, 5), rep(1.0, 5), rep(0.0, 5))
    cont.covs <- rmvn(n, mu, sigma = cov)
    bin.covs <- cbind(rbinom(n, size = 1, prob = 0.25), 
                      rbinom(n, size = 1, prob = 0.5), 
                      rbinom(n, size = 1, prob = 0.5),
                      rbinom(n, size = 1, prob = 0.75),
                      rbinom(n, size = 1, prob = 0.75))
    x <- cbind(bin.covs, cont.covs)

    ## nonlinear continuous  single
    mu4 <- 1 + 2 * x[ , 1] + 2 * x[ , 6] + sin( pi * x[ , 2] * x[ ,7] ) - 2 * exp( x[ , 3] * x[ , 5] ) +
           log( abs( cos ( pi / 2 * x[ , 8] ) ) ) - 1.8 * cos( x[ , 9]) + 3 * x[ , 3] * abs(x[ ,7]) ^ 1.5
    y4  <- rnorm(n, mu4 , sd = sd)

    mydat4      <- cbind(x, y4)
    dat4[k, , ] <- mydat4
  }
  return(list(  dat4 = dat4) )
}
n500_sd1 <- makedata(500, 500, 20, 5, 1)
n5000_sd1 <- makedata(500, 5000, 20, 5, 1)

\end{verbatim}

\subsection{Scenario 2}

\subsubsection{Part a}
\begin{verbatim}
library(mvnfast)
expit <- function(x) exp(x) / (1 + exp(x))
makedata <- function(nsim, n) {
  dat8 <- array(0, dim = c(nsim, n, 30 + 2) )
  for (k in 1:nsim) {
    p <- 30
    mu <- rep(0, p)
    sig <- matrix(0, nrow = p, ncol = p)
    diag(sig) <- 1
    rho <- 0.5
    for(i in 1:(p-1)) {
      for(j in (i+1):p) {
        kk <- j - i
        sig[i, j] <- sig[j, i] <- rho^kk
      }
    }
    x <- rmvn(n, mu, sigma = sig)
    prob.a <- expit(0.1 + 0.2 * x[,1] - sin(x[,3])/3 - 0.1 * x[,22])
    a <- rbinom(n, 1, prob.a)
    x <- cbind(a, x)
    ## nonlinear continuous -multi
    mu8 <- 1 + 2 * x[ , 1] - 1 * x[ , 1] * x[ , 2] + 2 * x[ , 2] + sin( pi * x[ , 22] * x[ ,7] ) - 1 * exp( x[ , 6]/5 * x[ , 5] ) +
           log( abs( cos ( pi / 2 * x[ , 8] ) ) ) - 1.8 * cos( x[ , 9]) + 0.2 * x[ , 11] * abs(x[ ,7]) ^ 1.5
    y8  <- rnorm(n, mu8 , sd = 1)
    mydat8      <- cbind(x, y8)
    dat8[k, , ] <- mydat8
  }
  return(list( dat8 = dat8) )
}
n500  <- makedata(500, 500)
n5000 <- makedata(500, 5000)
\end{verbatim}

\subsubsection{Part b}

\begin{verbatim}
library(mvnfast)
expit <- function(x) exp(x) / (1 + exp(x))
makedata <- function(nsim, n) {
  dat8 <- array(0, dim = c(nsim, n, 30 + 2) )
  for (k in 1:nsim) {
    p <- 30
    mu <- rep(0, p)
    sig <- matrix(0, nrow = p, ncol = p)
    diag(sig) <- 1
    rho <- 0.5
    for(i in 1:(p-1)) {
      for(j in (i+1):p) {
        kk <- j - i
        sig[i, j] <- sig[j, i] <- rho^kk
      }
    }
    x <- rmvn(n, mu, sigma = sig)
    prob.a <- expit(0.1 + 0.2 * x[,1] - 0.5*x[,2]-0.1*x[,1]*x[,2] + 0.3 *x[,4]+0.1*x[,5]+
                    0.7*x[,4]*x[,5] - 0.4*x[,11]*x[,22] - + 0.4*x[,10]^2*x[,15] - 0.1 * x[,22])
    a <- rbinom(n, 1, prob.a)
    x <- cbind(a, x)
    ## nonlinear continuous -multi
    mu8 <- 1 + 2 * x[ , 1] - 1 * x[ , 1] * x[ , 2] + 2 * x[ , 2] + 2 * sin( pi * x[ , 22] * x[ ,7] ) - 1 * exp( x[ , 6]/5 * x[ , 5] ) -
           log( abs( cos ( pi / 2 * x[ , 8] ) ) ) - 1.8 * cos( x[ , 9]) + 1.2 * x[ , 11] * abs(x[ ,7]) ^ 1.5
    mu8 <- 1 + 2 * x[ , 1] - 1 * x[ , 1] * x[ , 2] + 2 * x[ , 2] - 1 * x[ , 3] + 2 * x[ , 4] - 1.5 * x[ , 5] - 0.5 * x[ , 6] - 2*x[,7] +
           x[,4]^2 - x[,7]^2 + 2*x[,4]*x[,5] - 1 *x[,3]*x[,7] + 0.5 * x[,6]*x[,7] - 0.2*x[,3]*x[,4]*x[,5] + x[,7]*x[,9]*x[,10] - x[,8]*x[,22]*x[,25]*x[,26] +
           x[,11]*x[,14]*x[,15]*x[,27] - x[,25]*x[,26]^2*x[,11] + 3*x[,4]*x[,17]^2 - 3*x[,5]*x[,18]^2 + x[,4]*x[,5]*x[,10]*x[,15] -x[,4]*x[,5]*x[,10]*x[,15]^2+
           1.5*x[,11]*x[,22]
    y8  <- rnorm(n, mu8 , sd = 1)

    mydat8      <- cbind(x, y8)
    dat8[k, , ] <- mydat8
  }
  return(list(dat8 = dat8) )
}
n500  <- makedata(500, 500)
n5000 <- makedata(500, 5000)
\end{verbatim}

\subsection{Scenario 3}

\begin{verbatim}
library(mvnfast)
expit <- function(x) exp(x) / (1 + exp(x))
makedata <- function(nsim, n) {
  dat6 <- array(0, dim = c(nsim, n, 30 + 2) )
  for (k in 1:nsim) {
    p <- 30
    mu <- rep(0, p)
    sig <- matrix(0, nrow = p, ncol = p)
    diag(sig) <- 1
    rho <- 0.5
    for(i in 1:(p-1)) {
      for(j in (i+1):p) {
        kk <- j - i
        sig[i, j] <- sig[j, i] <- rho^kk
      }
    }
    x <- rmvn(n, mu, sigma = sig) 
    prob.a <- expit(0.1 + 0.2 * x[,1] - sin(x[,3])/3 - 0.1 * x[,22])
    a <- rbinom(n, 1, prob.a)
    x <- cbind(a, x)
     ## nonlinear binary - multi
     prob.y6 <- pnorm(0.1 + 0.3 * x[ , 1] - 0.1 * x[ , 1] * x[ , 2] + 0.1 * x[ , 2] - sin( pi / 4 * x[ , 22] * x[ , 7] ) + exp(x[,7] / 5)*x[,11]/4 -
                      0.12 * x[, 22] * x[ , 9] * x[ ,10] + 0.05*x[,8] * x[,10]*x[,11]^2)
     y6      <- rbinom(n, size = 1, prob = prob.y6)

    mydat6      <- cbind(x, y6)
    dat6[k, , ] <- mydat6
  }
  return(list( dat6 = dat6) )
}
n500  <- makedata(500, 500)
n5000 <- makedata(500, 5000)
\end{verbatim}

\subsection{Scenario 4}

\begin{verbatim}
library(mvnfast)
expit <- function(x) exp(x) / (1 + exp(x))
makedata <- function(nsim, n) {
  dat8 <- array(0, dim = c(nsim, n, 30 + 2) )
  for (k in 1:nsim) {
    p <- 30
    mu <- rep(0, p)
    sig <- matrix(0, nrow = p, ncol = p)
    diag(sig) <- 1
    rho <- 0.5
    for(i in 1:(p-1)) {
      for(j in (i+1):p) {
        kk <- j - i
        sig[i, j] <- sig[j, i] <- rho^kk
      }
    }
    x <- rmvn(n, mu, sigma = sig)
    prob.a <- expit(0.1 + 0.2 * x[,1] - sin(x[,3])/3 - 0.1 * x[,22])
    a <- rbinom(n, 1, prob.a)
    x <- cbind(a, x)
    ## nonlinear continuous -multi
    mu8 <- 1 + 2 * x[ , 1]  +  sin( pi * x[ , 22] * x[ ,7] ) - 1 * exp( x[ , 6]/5 * x[ , 5] ) +
           log( abs( cos ( pi / 2 * x[ , 8] ) ) ) - 1.8 * cos( x[ , 9]) + 0.2 * x[ , 11] * abs(x[ ,7]) ^ 1.5 + x[,2]*x[,3] - 0.5*x[,2]^2 - cos(x[,2])
    y8  <- rnorm(n, mu8 , sd = 1)
    mydat8      <- cbind(x, y8)
    dat8[k, , ] <- mydat8
  }
  return(list( dat8 = dat8) )
}
n500  <- makedata(500, 500)
n5000 <- makedata(500, 5000)
\end{verbatim}

\section{Additional results from data analysis}

\textbf{Trace plots for analysis}

\begin{figure}[H]
    \centering
    \includegraphics[width=0.6\textwidth]{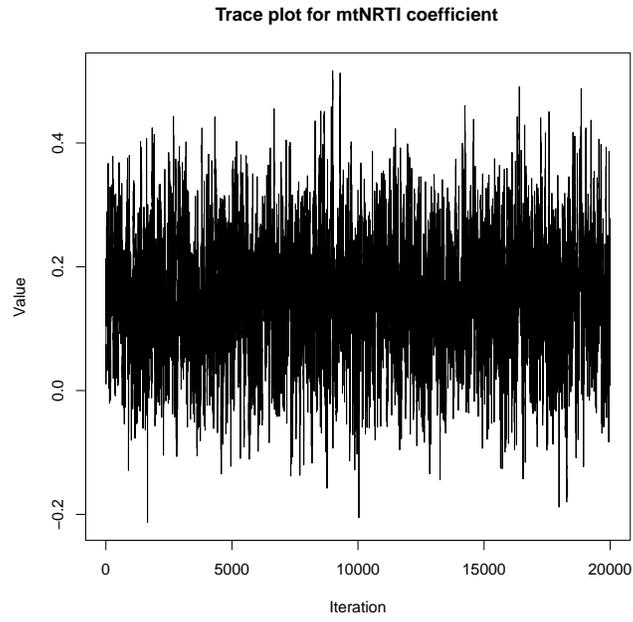}
    \caption{Trace plot for analysis with no effect modification.}
  \label{fig:supp1}
\end{figure}

\begin{figure}[H]
    \centering
    \includegraphics[width=0.7\textwidth]{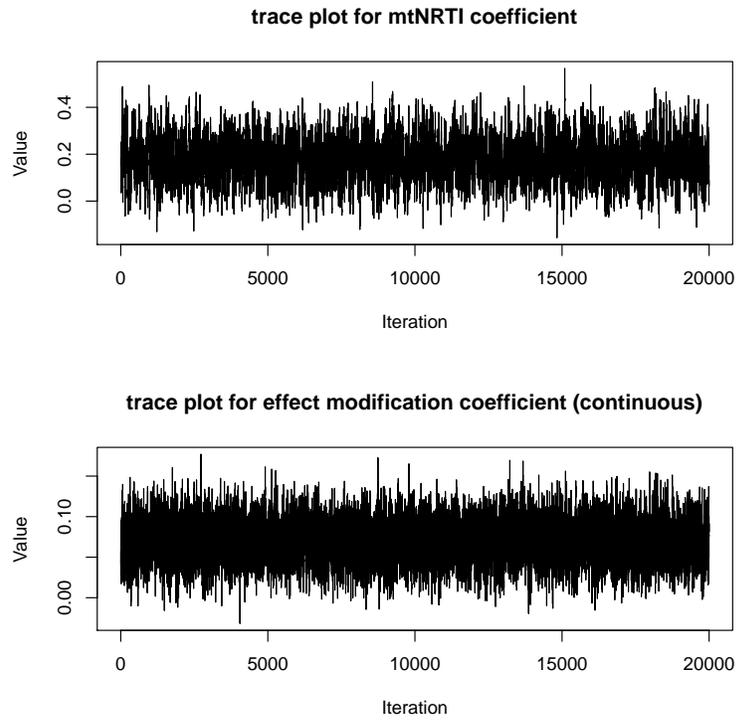}
    \caption{Trace plot for analysis with continuous effect modifier.}
  \label{fig:supp2}
\end{figure}

\begin{figure}[H]
    \centering
    \includegraphics[width=0.7\textwidth]{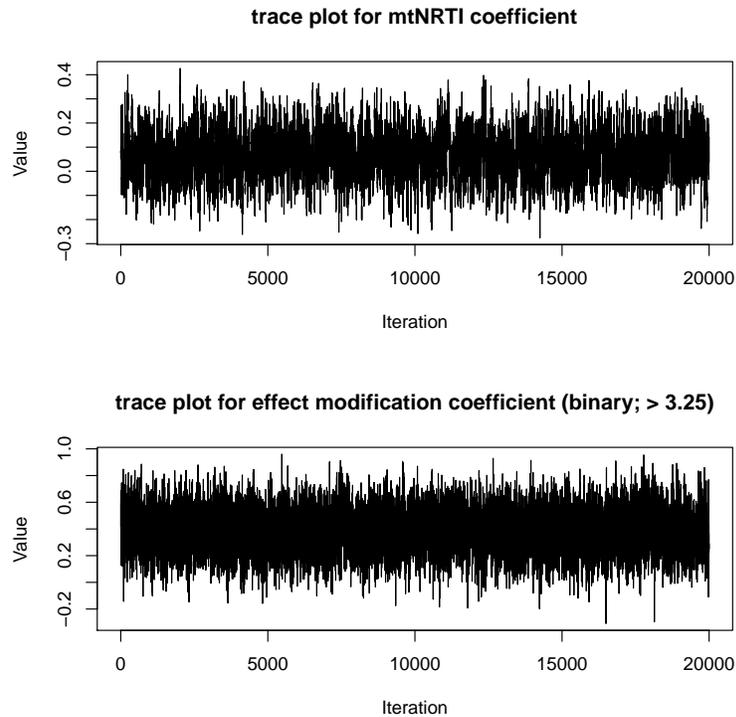}
    \caption{Trace plot for analysis with binary effect modifier.}
  \label{fig:supp3}
\end{figure}


\bibliographystyle{unsrtnat}
\bibliography{semibart}

\end{document}